\documentclass[prd,twocolumn,showpacs,preprintnumbers,amsmath,amssymb]{revtex4}

\usepackage{graphicx}
\usepackage{dcolumn}
\usepackage{bm}
\begin{document}
\title{Measurement of branching fractions for the inclusive
Cabibbo-favored $\bar{K}^{*0}(892)$ and Cabibbo-suppressed
$K^{*0}(892)$ decays of neutral and charged $D$ mesons}
\author{
M.~Ablikim$^{1}$,              J.~Z.~Bai$^{1}$,               Y.~Ban$^{11}$,
J.~G.~Bian$^{1}$,              X.~Cai$^{1}$,                  J.~F.~Chang$^{1}$,
H.~F.~Chen$^{16}$,             H.~S.~Chen$^{1}$,              H.~X.~Chen$^{1}$,
J.~C.~Chen$^{1}$
% ...
\\(BES Collaboration)}
\affiliation{
\begin{minipage}{145mm}
$^{1}$ Institute of High Energy Physics, Beijing 100049, P.R. China\\
% ...
$^{11}$ Peking University, Beijing 100871, P.R. China\\
% ...
$^{16}$ University of Science and Technology of China, Hefei 230026, P.R. China\\
% ...
\end{minipage}
}
\email{chenjc@mail.ihep.ac.cn}
\date{\today}

\begin{abstract}
{The branching fractions for the inclusive Cabibbo-favored 
$\bar{K}^{*0}$ and Cabibbo-suppressed $K^{*0}$ decays of $D$ mesons
are measured based on a data sample of 33 $pb^{-1}$
collected at and around the center-of-mass 
energy of 3.773 GeV with the BES-II detector at the BEPC collider.
The branching fractions for the decays
$D^{+(0)}\rightarrow \bar{K}^{*0}(892) X$ and
 $D^0 \rightarrow K^{*0}(892) X$ are determined to be
$BF(D^0\rightarrow \bar{K}^{*0} X)$ = $(8.7\pm 4.0\pm 1.2)\%$, 
$BF(D^+\rightarrow \bar{K}^{*0} X)$ = $(23.2\pm 4.5\pm3.0)\%$ and
 $BF(D^0 \rightarrow K^{*0} X)$ = $(2.8\pm 1.2\pm 
0.4)\%$.
An upper limit on the branching fraction at 90\% C.L. for the decay 
 $D^+ \rightarrow K^{*0}(892) X$ is set to be
$BF(D^+ \rightarrow K^{*0} X) < 6.6\%$. 
}
\end{abstract}


\maketitle

\section{Introduction}
Although $D$ mesons were found 29 years ago~\cite{D1,D2}, the study 
of charm meson decay properties is still an interesting field. 
Measurement of branching fractions for the $D$ meson decay 
modes containing 
$\bar{K}^{*0}$($K^{*0})$ in the final 
states can provide useful information about the relative strength
of the Cabibbo-favored and Cabibbo-suppressed $D$ decays. 
The total branching fractions for the exclusive $D$ decay modes  
 containing $\bar{K}^{*0}$ are summed to be $BF(D^0\rightarrow 
\bar{K}^{*0} X)=(8.1\pm0.8)\%$  and  
$BF(D^+\rightarrow
\bar{K}^{*0} X)=(23.1\pm2.0)\%$ ($X=any\ particles$) with the known 
branching fractions of 
 neutral and charged $D$ mesons~\cite{PDG}.
The measurement of branching fractions for the inclusive
Cabibbo-favored decay $D\rightarrow \bar{K}^{*0} X$
and Cabibbo-suppressed decay $D\rightarrow K^{*0} X$ can 
not only serve as an independent check on the sum of the branching 
fractions for the
exclusive $D$ decay modes containing $\bar{K}^{*0}$ meson in the final 
states, which indicates the need to search for new decay modes, but also 
provides valuable 
information in understanding the weak decay mechanism. The 
knowledge of the inclusive $D$ meson decay properties
will also help one to understand $B$ decays.

This Letter reports the measurement of branching fractions for 
the inclusive Cabibbo-favored decay $D\rightarrow \bar{K}^{*0} X$ and 
Cabibbo-suppressed decay $D\rightarrow K^{*0} X$ of
neutral and charged $D$ mesons using a double tag method 
described in 
Section III, based on an analysis 
of about 33 $pb^{-1}$ of
data collected with the upgraded 
Beijing Spectrometer (BES-II) in $e^+e^-$ annihilation at and around 
$\sqrt{s}=3.773$ GeV. 

\section{The Beijing Spectrometer}
BES is a conventional cylindrical magnetic detector ~\cite{BES}
operated at the Beijing Electron Positron Collider (BEPC) ~\cite{BEPC}.
BESII is the upgraded version of the BES detector ~\cite{BESII}.
A 12-layer vertex chamber (VC)
surrounding the beam pipe provides trigger information.
A forty-layer main drift chamber (MDC) located outside the VC 
performs trajectory and ionization energy loss ($dE/dx$) measurement
with a solid angle coverage of $85\%$ of $4\pi$ for charged tracks.
Momentum resolution of $\sigma_p/p = 1.7\%\sqrt{1+p^2}$ ($p$ in 
GeV/c)
and $dE/dx$ resolution of $8.5\%$ for Bhabha scattering electrons
are obtained for the
data taken at $\sqrt{s}=3.773$ GeV. An array of 48 scintillation 
counters
surrounds the MDC and measures the time of flight (TOF) of charged 
tracks
with a resolution of about 200 ps for the electrons.
Surrounding the TOF is a 12-radiation-length, lead-gas barrel shower counter
 (BSC) operated in limited streamer mode,
which measures the energies of electrons
and photons over $80\%$ of the total solid angle, and has an energy
resolution of $\sigma_E/E=0.22/\sqrt{E}$ ($E$ in GeV),
spatial resolutions of $\sigma_{\phi}=7.9$ mrad and $\sigma_Z=2.3$ cm for the
electrons. Outside of the BSC is a solenoidal magnet which provides a
0.4 T magnetic field in the central tracking region of the detector. Three
double-layer muon counters instrument the magnet flux return, and serve to
identify muons with momentum greater than 500 MeV/c. They cover $68\%$ 
of the
total solid angle with longitudinal (transverse) spatial resolution of 5 cm (3
cm). End-cap time-of-flight and shower counters extend coverage to the
forward and backward regions. A Monte Carlo package based on GEANT3 
has been developed for BESII detector simulation and 
comparisons with data show that the simulation is generally 
satisfactory~\cite{simbes}.

\section{Data Analysis}

\subsection{Event Selection}
For each event it is required that at least 2 (but no more than 10)
charged tracks are well reconstructed in the MDC with good helix fits.
All tracks, save those from $K_S^0$ decays, must originate from the
interaction region, which requires that for a charged track, the distance
of closest approach be less than 2 cm in the $xy$ plane, and less than 20 
cm in the $z$ direction.
For the $\pi^+$ and $\pi^-$
from the $K^0_S$ decay, the secondary vertex position is required to 
be less than 8 cm 
in the $xy$ plane and within $\pm 20$ cm in the $z$ direction to the 
primary interaction point.
In addition, in order to optimize the momentum resolution and
charged particle identification, a geometry cut
$\mid cos{\theta} \mid \leq 0.85$
 ($\theta$ is the polar angle of the track) is applied.
For the charged particle mass assignment, a combined 
confidence level calculated using the $dE/dx$ and TOF measurements
is required to be greater than $0.1\%$ for the pion hypothesis.
For the kaon identification, the confidence level for the kaon 
hypothesis 
is required to be greater than that for the pion hypothesis.

The $\pi^0$ is reconstructed through the decay 
$\pi^0 \rightarrow \gamma \gamma$. 
For the $\gamma$ from $\pi^0$ decay, the energy deposited in the 
BSC is required to be greater than 70 MeV;
the electromagnetic shower is required to start in the first 5 
readout layers; and the angle between the 
$\gamma$ and the nearest charged track is required to be greater than 
$22^{\circ}$.

\subsection{Singly Tagged $\bar{D}^0$ and $D^-$}
Around the center-of-mass energy of $3.773$ GeV, the
$\psi(3770)$ resonance is produced in electron-positron
 ($e^+e^-$) annihilation. The $\psi(3770)$ lies just above
  open charm pair production threshold and
decays predominantly into $D\bar{D}$ pairs.
If one $D$ meson is fully reconstructed (this is
called a singly tagged $\bar{D}$ event), the other $D$ meson must exist
on the recoil side. Throughout the paper, charge conjugation is implied.

For the analysis, singly tagged $\bar{D}$ events are reconstructed in
three hadronic $\bar{D}^0$ decay modes ($K^+\pi^-$, $K^+\pi^-\pi^-\pi^+$,
and $K^+\pi^-\pi^0$) and in nine $D^-$ decay 
modes ($K^+\pi^-\pi^-$, $K^0\pi^-$, $K^0K^-$, $K^+K^-\pi^-$,
$K^0\pi^-\pi^-\pi^+$, $K^0\pi^-\pi^0$, $K^+\pi^-\pi^-\pi^0$,
$K^+\pi^+\pi^-\pi^-\pi^-$ and $\pi^+\pi^-\pi^-$). 
The reconstruction for the singly tagged $\bar{D}$ events is the same 
 as that used in the previous works~\cite{Dsemi,Dsemi2}. 

In order to reduce the background and improve the momentum 
resolution,
a center-of-mass energy constraint 
kinematic fit (1C-fit) is imposed on each $mKn\pi$ (m=0,1,2;  
n=1,2,3,4) combination. 
For the singly tagged $\bar{D}$ decay modes with a neutral kaon 
or a neutral pion
in the daughter particles, 
an additional constraint kinematic fit for the 
$K_S^0 \rightarrow \pi^+\pi^-$ or 
$\pi^0 \rightarrow \gamma \gamma$ is also performed.
The kinematic fit probability $P(\chi^2)$ 
is required to be greater than $0.1\%$.
If more than one combination 
satisfies the criteria in an event, only the
combination with the largest $P(\chi^2)$ is retained. 

Fig.~\ref{d0_tag} and fig.~\ref{dp_tag} show the invariant mass 
spectra for 
$mKn\pi$ combinations in the singly tagged $\bar{D}^0$ and  
$D^-$ decay modes, which are calculated 
based on the track momentum vectors from the kinematic fit. 
A maximum likelihood fit to the mass spectrum with a Gaussian function 
for the $\bar{D}$ signal and a special background function 
(ARGUS background shape multiplied by a polynomial 
function)~\cite{Dsemi} to describe backgrounds yields 
a total number of $7033\pm193\pm316$ singly tagged $\bar{D}^0$ events and 
$5321\pm149\pm160$ singly tagged $D^-$ events (as well as the numbers 
 for the individual decay channels). The errors on the numbers of events are
statistical (first) and systematic (second) where the later is 
obtained by varying the parameterization of the background.

\begin{figure}
\includegraphics[width=8.5cm,height=6cm]
{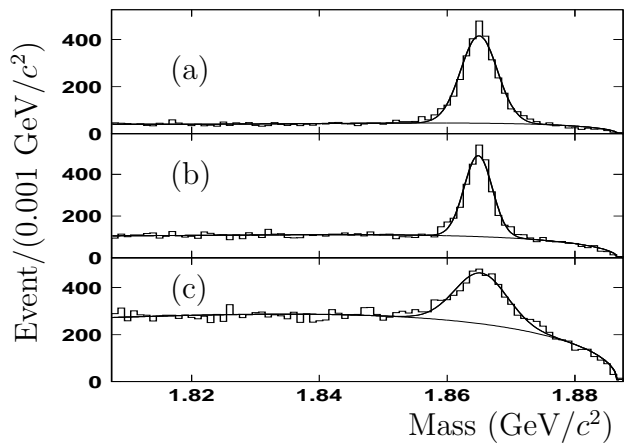}
\caption
{\label{d0_tag} Invariant mass spectra for 
$Kn\pi$ (n=1,2,3) combinations in 
the singly tagged $\bar{D}^0$ decay modes:
(a) $K^+\pi^-$,
(b) $K^+\pi^-\pi^-\pi^+$ and  
(c) $K^+\pi^-\pi^0$.
\label{d0_tag}}
   \begin{picture}(5,5)
      \put(-60,190){\large (a)}
      \put(-60,150){\large (b)}
      \put(-60,107){\large (c)}
      \put(30,55){\large Mass (GeV/$c^2$)}
      \put(-120,90){\rotatebox{90}{\large Event/(0.001 GeV/$c^2$)}}
   \end{picture}
\end{figure}

\begin{figure}
\includegraphics[width=8.5cm,height=6cm]
{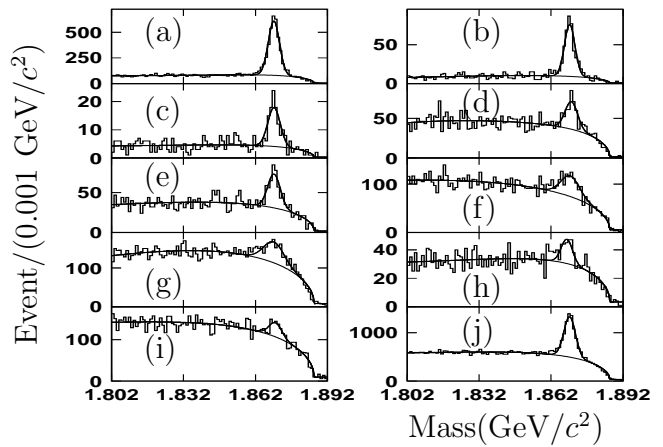}
\caption
{\label{dp_tag}  Invariant mass spectra for $mKn\pi$ 
(m=0,1,2,n=1,2,3,4) combinations
in the singly tagged $D^-$ decay modes:
 (a) $K^+\pi^-\pi^-$,
(b) $K^0\pi^-$, (c) $K^0K^-$, (d) $K^+K^-\pi^-$, 
(e) $K^0\pi^-\pi^-\pi^+$, (f) $K^0\pi^-\pi^0$, 
(g) $K^+\pi^-\pi^-\pi^0$, 
(h) $K^+\pi^+\pi^-\pi^-\pi^-$, 
(i) $\pi^+\pi^-\pi^-$ and (j) sum of nine modes.
\label{dp_tag}}
   \begin{picture}(5,5)
      \put(-70,225){\large (a)}
      \put(50,225){\large (b)}
      \put(-70,195){\large (c)}
      \put(50,198){\large (d)}
      \put(-70,170){\large (e)}
      \put(50,158){\large (f)}
      \put(-70,130){\large (g)}
      \put(50,128){\large (h)}
      \put(-70,105){\large (i)}
      \put(50,112){\large (j)}
      \put(30,75){\large Mass(GeV/$c^2$)}
      \put(-120,110){\rotatebox{90}{\large Event/(0.001 
GeV/$c^2$)}}
   \end{picture}
\end{figure}

\subsection{Doubly Tagged Events}
\subsubsection{Inclusive decay $D\rightarrow \bar{K}^{*0}(K^{*0}) X$}
The inclusive decay
$D\rightarrow \bar{K}^{*0}X$ is reconstructed on the recoil side of 
the singly tagged
$\bar{D}$, 
where the $\bar{K}^{*0}$ is reconstructed through
its decay to $K^-\pi^+$. 
The regions within a $\pm3\sigma_{M_D}$ window around the 
fitted $\bar{D}$ masses in the invariant mass spectra for $mKn\pi$ 
combinations, as shown in 
fig.~\ref{d0_tag} and fig.~\ref{dp_tag}, 
are defined as the singly tagged $\bar{D}$ signal regions 
($\bar{D}\ tag\ region$), 
where $\sigma_{M_D}$ are the standard deviations of the mass spectra.
Those outside a $\pm 4\sigma_{M_D}$ window around the 
fitted $\bar{D}$ masses are taken as the sideband background
control regions ($\bar{D}\ sideband$). In the estimation of the 
number of background events in the $\bar{D}$ tag region, the number
 of the events in the $\bar{D}$ sideband is normalized to 
the  area of the fitted background in the $\bar{D}$ tag region.

 Due to particle misidentification 
and random combination, the invariant mass for $K^-\pi^+$ combination 
 could enter into mass spectra more than once per event.
To avoid this problem, only the $K^-\pi^+$ combination with 
the maximum product of the confidence levels for the $K$  
hypothesis and $\pi$ hypothesis is retained in an event.

\begin{figure}[hbt]
\includegraphics[width=7.5cm,height=6cm]
{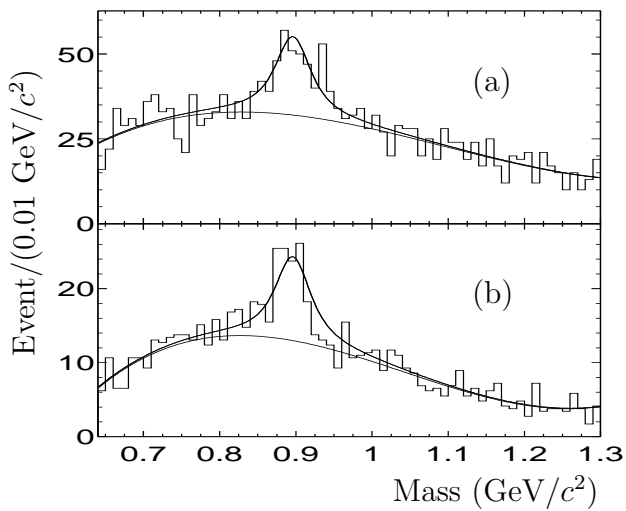}
\vspace{0.5cm}
   \begin{picture}(6,5)
      \put(-60,140){\large (a)}
      \put(-60,60){\large (b)}
      \put(-235,40){\rotatebox{90}{\large Event/(0.01 GeV/$c^2$)}}
      \put(-90,-15){\large Mass (GeV/$c^2$)}
   \end{picture}
\caption{\label{d0_kpi}
Invariant mass spectra for $ K^-\pi^+$ combinations observed
on the recoil side of the $\bar{D}^0$ tags
for study of the Cabibbo-favored decay $D^0\rightarrow \bar{K}^{*0} 
X$: 
(a) the mass spectrum for the events with a $\bar{D}^0$ tag, 
(b) the normalized mass spectrum for the $\bar{D}^0$ sideband events.
}
\end{figure}

Invariant mass spectra for $ K^-\pi^+$ combinations observed
on the recoil side of the $\bar{D}^0$ tags (the $mKn\pi$ 
combinations) are 
shown in fig.~\ref{d0_kpi}  for study of the 
Cabibbo-favored decay $D^0\rightarrow \bar{K}^{*0} X$, 
where (a) shows
the mass spectrum for the events with a tagged $\bar{D}^0$, 
and (b) shows the normalized mass spectrum 
of the $\bar{D}^0$ sideband events. By fitting the invariant mass spectra for
$K^-\pi^+$ combinations 
with a Gaussian function for the $\bar{K}^{*0}$ signal and a polynomial 
to describe background, the numbers of the observed $\bar{K}^{*0}$ events are 
obtained to be $188.5\pm37.3$ and $92.6\pm 23.5$ for the tagged and 
sideband events, respectively. 
In the fit, 
the mass and width of $\bar{K}^{*0}$ are fixed at
0.8961 GeV/$c^2$ and 0.0507 GeV/$c^2$ quoted from PDG~\cite{PDG}, 
and the detector resolution is set to be 0.0078 GeV/$c^2$, which is 
obtained by 
fitting the invariant mass spectrum for $K^-\pi^+$ combinations from both
$D^0$ and $D^+$ decays. 
After subtracting the number of the background events, we obtain
$95.9\pm 44.1$ signal events for the $D^0\rightarrow 
\bar{K}^{*0}X$ decay.

\begin{figure}[hbt]
\includegraphics[width=7.5cm,height=6cm]{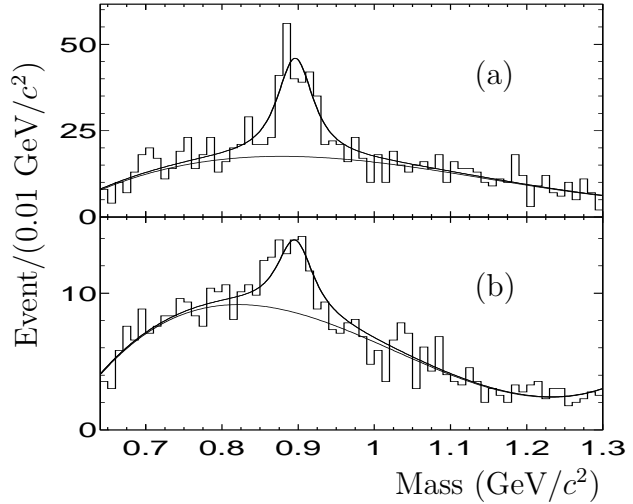}
\vspace{0.5cm}
   \begin{picture}(5,5)
      \put(-60,140){\large (a)}
      \put(-60,60){\large (b)}
      \put(-235,40){\rotatebox{90}{\large Event/(0.01 GeV/$c^2$)}}
      \put(-90,-15){\large Mass (GeV/$c^2$)}
   \end{picture}
\caption{\label{dp_kpi} 
Invariant mass spectra for $ K^-\pi^+$ combinations observed
on the recoil side of the $D^-$ tags
for study of the Cabibbo-favored decay $D^+\rightarrow 
\bar{K}^{*0} X$: 
(a) the mass spectrum for the events with a $D^-$ tag, 
(b) the normalized mass spectrum for the $D^-$ sideband events.
}
\end{figure}

\begin{figure}[hbt]
\includegraphics[width=7.5cm,height=6cm]
{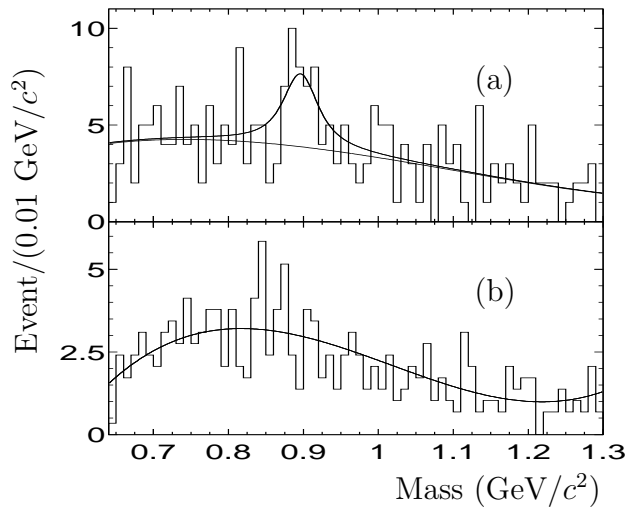}
\vspace{0.5cm}
   \begin{picture}(5,5)
      \put(-60,140){\large (a)}
      \put(-60,60){\large (b)}
      \put(-235,40){\rotatebox{90}{\large Event/(0.01 GeV/$c^2$)}}
      \put(-90,-15){\large Mass (GeV/$c^2$)}
   \end{picture}
\caption{\label{d0_kpi_cpd} Invariant mass spectra for
$K^+\pi^-$ combinations observed on the recoil side of the 
$\bar{D}^0$ tags for study of the Cabibbo-suppressed decay 
$D^0\rightarrow K^{*0}X$: 
(a) the mass spectrum for the events with a $\bar{D}^0$ tag, 
(b) the normalized mass spectrum for the $\bar{D}^0$ sideband events.
 }
\end{figure}

\begin{figure}
\includegraphics[width=7.5cm,height=6cm]
{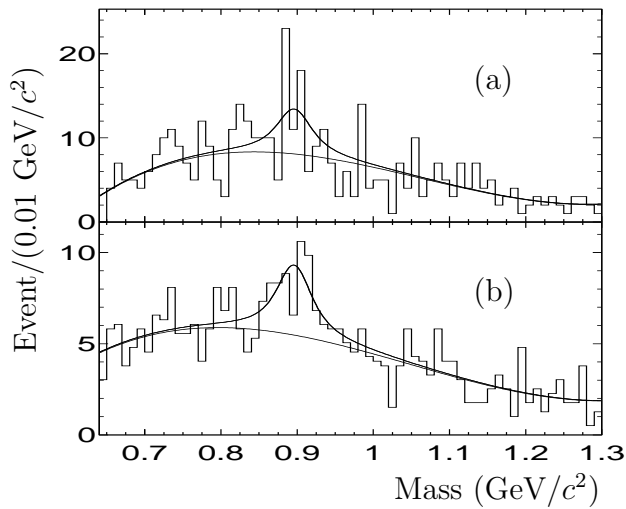}
\vspace{0.5cm}
   \begin{picture}(5,5)
      \put(-60,140){\large (a)}
      \put(-60,60){\large (b)}
      \put(-235,40){\rotatebox{90}{\large Event/(0.01 GeV/$c^2$)}}
      \put(-90,-15){\large Mass (GeV/$c^2$)}
   \end{picture}
\caption{\label{dp_kpi_cpd} Invariant mass spectra for
$K^+\pi^-$ combinations observed on the recoil side of the $D^-$ tags 
for study of the Cabibbo-suppressed decay 
$D^+\rightarrow K^{*0} X$: 
(a) the mass spectrum for the events with a $D^-$ tag,  
(b) the normalized mass spectrum for the $D^-$ sideband events.}
\end{figure}

Similarly, fig.~\ref{dp_kpi} shows the invariant mass
spectra for $ K^-\pi^+$ combinations selected on the recoil side of 
$D^-$ tags to study the
Cabibbo-favored decay $D^+\rightarrow \bar{K}^{*0} X$.
Fig.~\ref{d0_kpi_cpd} and fig.~\ref{dp_kpi_cpd} show the invariant mass
spectra for $ K^+\pi^-$ combinations on the recoil side of 
$\bar{D}^0$ and $D^-$ tags to study the 
Cabibbo-suppressed decays  $D^0\rightarrow K^{*0} X$ and
$D^+\rightarrow K^{*0} X$. With the same analysis procedure 
as above, the numbers of the observed $\bar{K}^{*0}/K^{*0}$ 
events are obtained (Table~\ref{table1}).  
After subtracting the number of the background events, 
we obtain $189.1\pm36.0$, 
$30.8\pm13.2$ and $12.3\pm23.3$ signal events for $D^+\rightarrow 
\bar{K}^{*0}X$,  $D^0\rightarrow 
K^{*0}X$ and  $D^+\rightarrow K^{*0}X$ decays, respectively.

\begin{table}
\begin{center}
\caption{Number of $\bar{K}^{*0}/K^{*0}$ events observed on 
the recoil side of the 
 $\bar{D}$ tags, where $N$ and $N_b$ are the number of 
$\bar{K}^{*0}/K^{*0}$ events observed from the events in which 
the invariant masses of the $mKn\pi$ combinations are in 
 the $\bar{D}$ signal region and  in
the $\bar{D}$ sideband region, respectively.  n is
 the number of the signal events for $D$ decays.
}
\vspace{0.2cm}
\begin{tabular}{|c|c|c|c|}
\hline
Decay Mode &$N$  & $N_b$ & n \\
\hline
$D^0\rightarrow \bar{K}^{*0}X$&$188.5\pm37.3$ & $92.6\pm23.5$
&$95.9\pm44.1$
\\
$D^0\rightarrow K^{*0}X$&$30.8\pm13.2$ &$0.0\pm0.1$ &$30.8\pm13.2$
\\
\hline
$D^+\rightarrow \bar{K}^{*0}X$&$232.5\pm30.9$ & $43.4\pm18.4$
&$189.1\pm36.0$
\\
$D^+\rightarrow K^{*0}X$&$43.7\pm17.6$ & $31.4\pm15.2$
&$12.3\pm23.3$ \\
\hline
\end{tabular}
\label{table1}
\end{center}
\end{table}

\subsubsection{Efficiencies for $D\rightarrow  \bar{K}^{*0} (K^{*0}) X$}
The efficiencies for reconstruction of the inclusive $\bar{K}^{*0}$ 
decays of $D$ mesons are estimated with the Monte Carlo simulation. 
 The Monte Carlo events are generated as $e^+e^-\rightarrow D\bar{D}$, 
where $\bar{D}$ decays into the singly tagged $\bar{D}$ modes and 
$D$ decays into $\bar{K}^{*0}X$.
The particle trajectories are simulated with the GEANT3 based Monte 
Carlo simulation package of the BESII detector~\cite{simbes}. 
The average efficiencies are obtained by weighting
the branching fractions of $D$ meson decays quoted from PDG~\cite{PDG}
and the numbers of the singly tagged $\bar{D}$ events.
The efficiencies are  $0.1575\pm0.0015$ for the decay $D^0 
\rightarrow \bar{K}^{*0} (K^{*0}) X$ and $0.1529\pm0.0021$ for 
the decay $D^+\rightarrow  \bar{K}^{*0} (K^{*0}) X$.

\section{Results}

With the numbers of the observed signal events for the decay
$D \rightarrow \bar{K}^{*0} X$, 
the numbers of the singly tagged $\bar{D}$ mesons
and the reconstruction efficiencies,
the branching fractions 
for the Cabibbo-favored decay $D \rightarrow \bar{K}^{*0} X$ are 
determined to be
\begin{equation}
BF(D^0 \rightarrow \bar{K}^{*0} X)= (8.7\pm 4.0 \pm 1.2)\%
\label{eq2}
\end{equation}
and
\begin{equation}
BF(D^+ \rightarrow \bar{K}^{*0} X)= (23.2 \pm 4.5 \pm 3.0)\% .
\label{eq3}
\end{equation}
These results are consistent with those measured by the BES collaboration 
based on 
the data taken at $\sqrt{s}=4.03$ GeV~\cite{BES1}.

For the Cabibbo-suppressed decay $D^0 
\rightarrow K^{*0} X$ the branching fraction is obtained to be
\begin{equation}
BF(D^0 \rightarrow K^{*0}X)= (2.8\pm 1.2 \pm 0.4)\%.
\label{eq4_1}
\end{equation}
An upper limit on the branching fraction at 90\% C.L. for
the Cabibbo-suppressed decay $D^+ \rightarrow K^{*0}X$ is set
 to be
\begin{equation}
BF(D^+ \rightarrow K^{*0}X) < 6.6\%,
\label{eq6}
\end{equation}
which includes the systematic uncertainty.

If we treat the observed events as signal events, the branching 
fraction for the Cabibbo-suppressed  decay $D^+ \rightarrow 
K^{*0} X$ is 
\begin{equation}
BF(D^+ \rightarrow K^{*0}X)= (1.5_{-1.0}^{+2.9} \pm 0.2)\%.
\label{eq4}
\end{equation}

In the measured branching fractions,
the first error is statistical and second systematic.
The systematic error arises from the uncertainties in particle 
identification ($\sim1.0\%$)~\cite{Dsemi}, in tracking  
($2.0\%$ per track), 
in the numbers of the singly tagged $\bar{D}^0$ ($\sim4.5\%$) and 
 $D^-$ ($\sim3.0\%$)~\cite{Dsemi}, 
in background parameterization ($\sim12\%$), and in 
Monte Carlo statistics ($0.16\%$ for $D^0$ and $0.31\%$ for 
$D^+$).
These uncertainties are added
in quadrature to obtain the total systematic error, which is
$13.5\%$ for the $D^0$ decay and $13.0\%$ for the $D^+$ decay. 

\section{Summary}
Based on an analysis of about 33 $pb^{-1}$ of data collected with
the BES-II detector at the BEPC collider, we measured branching 
fractions for the inclusive Cabibbo-favored decay
$D\rightarrow \bar{K}^{*0} X$ and 
Cabibbo-suppressed decay $D\rightarrow K^{*0} X$. The results are  
$BF(D^0 \rightarrow \bar{K}^{*0} X)= (8.7\pm 4.0 \pm 1.2)\%$, 
$BF(D^+ \rightarrow \bar{K}^{*0} X)= (23.2 \pm 4.5 \pm 3.0)\%$
 and $BF(D^0 \rightarrow K^{*0}X)= (2.8\pm 1.2 \pm 0.4)\%$.
We set an upper limit on the branching fraction at 90\% C.L. for 
the decay $D^+ \rightarrow K^{*0}X$
 to be $BF(D^+ \rightarrow K^{*0}X) < 6.6\%$.

\section{Acknowledgments}
   The BES collaboration thanks the staff of BEPC for their diligent efforts.
This work is supported in part by the National Natural Science Foundation
of China under contracts
Nos. 10491300, 10225524, 10225525, the Chinese Academy
of Sciences under contract No. KJ 95T-03, the 100 Talents Program of CAS
under Contract Nos. U-11, U-24, U-25, and the Knowledge Innovation Project
of
CAS under Contract Nos. U-602, U-34 (IHEP); and by the
National Natural Science
Foundation of China under Contract No. 10175060 (USTC), and No. 10225522
(Tsinghua University).

\end{document}